# The View from the Other Side: The Border Between Controversial Speech and Harassment on Kotaku in Action

SHAGUN JHAVER, Georgia Institute of Technology
LARRY CHAN, Georgia Institute of Technology
AMY BRUCKMAN, Georgia Institute of Technology

In this paper, we use mixed methods to study a controversial Internet site: The Kotaku in Action (KiA) subreddit. Members of KiA are part of GamerGate, a distributed social movement. We present an emic account of what takes place on KiA: who are they, what are their goals and beliefs, and what rules do they follow. Members of GamerGate in general and KiA in particular have often been accused of harassment. However, KiA site policies explicitly prohibit such behavior, and members insist that they have been falsely accused. Underlying the controversy over whether KiA supports harassment is a complex disagreement about what "harassment" is, and where to draw the line between freedom of expression and censorship. We propose a model that characterizes perceptions of controversial speech, dividing it into four categories: criticism, insult, public shaming, and harassment. We also discuss design solutions that address the challenges of moderating harassment without impinging on free speech, and communicating across different ideologies.

## Contents

## Introduction

In a 2015 podcast on the radio show "This American Life," writer Lindy West interviews a man who viciously harassed her over the Internet (West (Speaker), 2015). In the process, her harasser comes to recognize her humanity and apologizes for his behavior. In return, West comes to understand her harasser as a person, and the personal challenges that explain (though don't excuse) his behavior. Inspired by West's experiences, we decided to try to understand online harassment in a more nuanced way by talking directly to harassers.

Specifically, we chose to study Kotaku in Action (KiA), a discussion forum for members of GamerGate. GamerGate is an online social movement portrayed in the popular press as a misogynistic hate group. A Washington Post article described GamerGate as "the freewheeling catastrophe/social movement/misdirected lynchmob that has, since August, trapped wide swaths of the Internet in its clutches, [and] has still — inexplicably! — not burned itself out." (Dewey, 2014). What is behind that portrayal? In this research we wanted to understand: Who are these people and how do they see their group and their activities? How do features of their sociotechnical system and the popular perceptions of their activities influence their behavior? What can we learn about controversial speech and harassment by considering their perspectives?

As researchers, we value listening. Individuals usually have more complex views than stereotypes predict. It is valuable to listen to the point of view of *anyone*, and listening does not imply either accepting or rejecting that person's view of the world. Inspired by Coleman's studies of the group Anonymous (Coleman, 2014) and Phillips' work on trolls (Phillips, 2015), we see value in trying to understand how members of KiA understand their commitments and practices. *Nothing in this paper should be construed to either support or attack either side of the GamerGate controversy.* The point of this work is to use this rich context to develop a more nuanced understanding of the complex boundary between free speech and harassment.

We begin with a discussion of online harassment and freedom of speech. We build on this discussion later to unpack our findings about KiA and describe different perceptions of controversial speech. Next, we present our methods of data collection and analysis. We divide our findings into two sections: a portrait of KiA (its members and activities), and members' views on free speech and harassment. Finally, we discuss broader implications for theory and design.

## Background

### *Online Harassment*

The problems of online harassment and digital hate crimes have grown increasingly salient in the past few years. A recent Data & Society Research Institute report based on a nationally representative survey found that 47 percent of Internet users have experienced online harassment and 72 percent of Internet users have witnessed at least one harassing behavior online (Lenhart et al., 2016).

Online harassment can have a deeply negative impact on its victims. They can suffer from trauma, anxiety, depression and other emotional problems. In some cases, they may even commit suicide (Ashktorab and Vitak, 2016). Online harassment can also damage an individual's reputation and safety because of the persistence and searchability of digital media (Duggan, 2014; Citron, 2014). In her book "Hate Crimes in CyberSpace," Danielle Citron highlights the profound emotional and physical suffering experienced by many cyber harassment victims who are often females or minorities (Citron, 2014). She proposes that achieving equality in digital networks should be the next stage of the women's rights movement. However, in the United States, Internet speech is protected under the First Amendment, making the ability to regulate hate speech problematic (Marwick and Miller, 2014).

One reason why combating online harassment is challenging is that it is often difficult to reach a consensus on which action crosses a line and which doesn't. Although online harassment has attracted a lot of media attention and research interest in recent years, there is no standard definition of what online harassment entails (Jhaver et al., 2018). This makes it difficult for social media platforms to battle abusive behaviors because they don't want to be seen as censoring what people can say online. A discussion of how to efficiently address online harassment requires that we start having conversations about what exactly it is.

Some researchers have attempted to conceptualize online harassment. Lwin et al. define online harassment as "rude, threatening or offensive content directed at others by friends or strangers, through the use of information communications technology" (Lwin et al., 2012). Ybarra et al. define it as "an intentional and overt act of aggression toward another person online" [Ybarra and Mitchell 2004]. However, what should be considered "offensive content" or an "act of aggression" can be subjective. Everyone agrees that posts of death threats and rape threats are abusive and should be regulated. But beyond such posts, where do we draw the line? How do we distinguish someone deliberately trying to harm another user from someone passionately disagreeing with that user? We use the findings from this study to explore these questions.

### *Freedom of Expression and its Limits*

As we will discuss later in this section, the Internet creates jurisdictional challenges for regulating content, because users from different countries with different laws on what speech is acceptable can post on the same forums. In the United States, the First Amendment states that government can not abridge citizens' freedom of speech, but this does not apply to private internet companies. We first present an overview of free speech and its limits before we unpack these issues in this section.

The problem of online harassment creates a quandary for strong advocates of freedom of speech. When does one person's right to free speech start to impinge on another person's right to be free from harassment? Many Americans consider free speech a universal and self-evident right. However, free speech is both a cultural norm and a legal construct, and its interpretations vary widely across different countries.

The notions of free speech in the US are unique, even among democratic countries. Many Western democracies like The United Kingdom, France, Germany and The Netherlands have laws against hate speech or speech that incites racial hatred. In contrast, the US protects free speech to the extent that it even protects racially and religiously offensive material (Zoller, 2009; Oetheimer, 2009). Most of Western Europe has banned holocaust denial, but there are no laws against it in the US. Furthermore, these laws affect but not completely determine how free speech is interpreted in different countries. We discuss later how social norms determine what is acceptable in different contexts.

Although Reddit is a platform with international reach, it is headquartered in the United States, and its approach to free speech is grounded in American values. In the United States, the First Amendment to the Constitution prohibits the government from censoring what citizens can say: "Congress shall make no law…abridging the freedom of speech, or of the press." Many contemporary First Amendment scholars generally believe that the risks of limiting expression always outweigh the risks of giving free rein to speech, and they advocate protecting all utterances and publications without discrimination (Canavan, 1984). They view freedom of expression as an absolute, overriding end in itself and warn of the dangers of the "slippery slope" – a current acceptable change to the status quo regarding speech can lead to some intolerable future limitations on speech if speech prohibition is introduced, they argue.

Some scholars have challenged this absolutist interpretation of the First Amendment. For example, Francis Canavan asserts that the purpose of the First Amendment is to protect and facilitate the achievement of rational ends by communication among ordinarily intelligent people (Canavan, 1984). He argues that when speech does not contribute to this purpose, it should not be protected.

*Freedom of Speech on Internet Platforms*

In the early days of the Internet, many academics and entrepreneurs of Silicon Valley shared a vision of freedom on the Internet. They envisioned the Internet as a forum for users to connect with one another without regard to race, gender, age or geography. They opposed legislation that would increase government regulation of the web and expected that the web would be adequately governed by the users themselves (Wortham, 2017).

However, this optimistic vision underestimated what would happen as the Internet grew. Unfortunately, with the growing opportunities for users to connect with one another online through social network sites, the Internet has also attracted a lot of disturbing behaviors. Such behaviors have been found to be pervasive and difficult to regulate on many sites. One reason for this failure is that the founding values of freedom on the Internet are so ingrained that many users and critics dislike regulation of any content on the web. Many Internet companies also embrace a libertarian view and try to avoid regulation in technology (Wortham, 2017).

A number of scholars have urged caution about the consequences of allowing extreme forms of speech. For example, Nancy Kim argues that applying the First Amendment analysis to the free speech versus online harassment debate without recognizing the ways in which online communication differs from offline communication fails to address many of the harms of online harassment (Kim, 2009). When harassment is conducted online, it can have more serious consequences because it is easier for digital information to spread faster and more widely. Danielle Citron argues that an absolutist devotion to free speech "needs to be viewed in light of the important interests that online harassment jeopardizes (Citron, 2014)." She writes that many harassers silence their victims and "we need to account for the full breadth of expression imperiled" when we evaluate the risks to expression that exist in our efforts to regulate online abuse.

Some critics argue that although hate speech is legal in the US, Internet platforms do not need to allow it. Laws about freedom of speech in the US control what the government can do—not private individuals. A corporate platform is more akin to a private party or club and it is legal to set rules for what may and may not be said there. However, other scholars have criticized the privatization of Internet content control and lament the absence of places on the Internet where free speech is constitutionally protected (Nunziato, 2005; Schesser, 2006). DeNardis has argued that since globalization and technological change have reduced the capacity of sovereign nation states to control information flows, Internet governance has now become the central front of freedom of expression (DeNardis, 2012; DeNardis and Hackl, 2016). In the same vein, Nunziato has noted the compelling private interests to provide public online spaces and advocated for legislatures to faithfully translate First Amendment values in cyberspace in order to make them meaningful in the technological age (Nunziato, 2005).

The overview of free speech and online harassment we present above is brief and far from exhaustive. However, it starts to reveal some of the challenges of regulating abusive behaviors and distinguishing disagreements from online harassment. We argue that online activities that are intentionally aggressive towards other individuals and threaten them should be considered as instances of online harassment. However, it is challenging to regulate actions which may be considered impolite and offensive by some individuals but which are civil and conducive to fostering democratic goals.

We use this perspective to present an emic account of the subreddit Kotaku in Action. As we will see, members see their group and their practices quite differently from their typical portrayal in both the popular press and academic literature. In seeing things from KiA members' point of view, we develop more broadly relevant insights into the boundary between freedom of expression and harassment.

### *Political Correctness*

As we will see, a central theme of discussion on KiA is expressing anger about the spread of "political correctness." What is "political correctness"? The term first emerged in Communist terminology in the 1930's to refer to the orthodox Communist party political line. It became part of the modern lexicon as a result of the public debates on university campuses in the United States between the liberal-left and the conservatives in the late 1980s. Geoffrey Hughes describes political correctness as a "complex, discontinuous and protean" sociolinguistic phenomenon that has "ramified from its initial concerns with education and the curriculum into numerous agendas, reforms, and issues concerning race, culture, gender, disability, the environment, and animal rights" (Hughes, 2010).

In its modern usage, the term "political correctness" or "PC" is used to describe words and actions that are chosen to avoid disparaging, offending or disadvantaging people, especially those belonging to oppressed groups in society. Many people, particularly conservatives, consider these actions as excessive and use the term "PC" as a pejorative. They see it as violating the principle of free speech, and consider it a means used by liberals to enforce conformity and stifle debates.

Commenters on the left complain that the forces against political correctness have used exaggeration and distortion to create the mythology of PC in order to divert attention and silence discussions about questions of equality and discrimination. They argue that the use of the term narrows the bounds of permissible political debate, and discourages cultural activism (Wilson, 1995).

*In this research, we explore how the KiA community views "PC culture," and how their perceptions shape their ideology and behavior. Our findings contribute to the broader conversation of how political correctness is affecting our culture and institutions.*Research Questions

In this research, we ask: Who are the people who participate in KiA? What are their shared beliefs and goals? How do the features of their sociotechnical system shape their emergent culture, and their tendency to be in conflict with other online groups? Do they see themselves as harassing others? What do they think "harassment" is? Most importantly, what can we learn about the nature of harassment by understanding nuances of its manifestation in this context?

## Gamergate and Kotaku in Action

### GamerGate

GamerGate is a distributed social movement that emerged in August 2014 (Kain, 2014). It began with a series of controversial events surrounding game developer Zoe Quinn, who was harassed as a result (Jason, 2015). This sparked a broader movement which initially focused on "ethics in game journalism," but quickly expanded to address broader issues of censorship, negative stereotyping of nerd culture, "social justice warrior" (SJW) [1] ideology, and perceived excesses of political correctness (Glasgow, 2015). Mortensen provides a detailed account of the progression of events that helped GamerGate gain popular attention (Mortensen, 2016).

Opponents of GamerGate have experienced doxing (revealing someone's personal information online), death threats, rape threats, and SWATing (tricking the police into raiding someone's home). Many who support GamerGate disavow these tactics (Glasgow, 2015), and insist that the perpetrators do not represent them. GamerGate does not have a defined membership or official leaders, so it is difficult to state whether "GamerGate" committed any particular act. Moreover, many GamerGate supporters claim that they also experienced doxing and harassment.

---

[1] Among GamerGate supporters, "social justice warrior" or SJW is a pejorative term for someone who, they claim, repeatedly makes shallow arguments about social justice for the purpose of raising their own personal reputation (Know Your Meme, 2016).

Like many contemporary online phenomena, GamerGate is not something that happens on one online site, but on a collection of sites with complex interactions across them (Gonzales et al., 2015). Supporters use the hashtag #GamerGate on Twitter, and websites like Reddit, 8chan, Voat and Tumblr for communication and collaboration (Wiki, 2016).

Although we initially set out to study GamerGate more generally, it rapidly became apparent that this was an impossibly huge task. To focus our efforts, we chose one online community dedicated to discussion of GamerGate issues: the Kotaku in Action (KiA) subreddit (Kotaku In Action, 2016).

### Kotaku in Action

Reddit is a social news website where users can submit content, either as a text post or as a link to another webpage. KiA describes itself as "the main hub for GamerGate discussion on Reddit." It is titled "Kotaku in Action" because at the time of its creation, it was dedicated to satirize Kotaku (a news and opinion site about games) for its alleged unethical journalistic practices. The sidebar of KiA declares its mission as: "KotakuInAction is a platform for open discussion of the issues where gaming, nerd culture, the Internet, and media collide."

As of December 8, 2017, KiA has 88,496 subscribers, and hundreds of active users at any given time. Its discussion board continues to remain active with dozens of new submissions every day.

KiA is just one of many sites for discussions on GamerGate. In fact, Reddit itself hosts a number of other subreddits for discussions related to GamerGate like "SocialJusticeInAction" and "GGDiscussion." A popular multi-reddit (a Reddit feature that enables combining and subscribing to several subreddits together) called "KiA HUB" collates submissions from ten such subreddits. There also exists subreddits like "r/GamerGhazi," that are devoted to anti-GamerGate discussions.

As Treré points out, "restricting the focus to only one of the many online technological manifestations of social movements risks overlooking important aspects such as the role and evolution of different platforms within a movement" (Treré, 2012). Therefore, we must note here that to fully understand the dynamics of GamerGate, future research must also consider Twitter as well as comparatively obscure sites like 8chan and Voat used by GamerGate supporters and opponents.

KiA is generally one of the mildest GamerGate forums, with less controversial speech than discussions of the topic on other sites like Twitter, 8chan or Voat. Although the popular press portrays GamerGate as a movement of misogynist Internet trolls (Allaway, 2014), we found that KiA members do not view themselves as such. Values our participants embrace include a strong support for freedom of speech, the view that political correctness has gone too far, the idea that white men are discriminated against in today's society, and a belief that the quality of journalism is in decline and the mainstream press too often blindly follows the values of PC (politically correct) culture.

## Methods

We begin this section by briefly discussing how we created rapport with the KiA community. Jennifer A. Rode writes that "discussions of rapport, even the cultural bumbling of getting it wrong, is critical to the ethnographic enterprise" (Rode, 2011). In particular, KiA was suspicious of outsiders studying it because it felt betrayed by previous occasions of journalists misrepresenting the community after interacting with its users. We hope that our discussion of rapport building contributes to the reader's understanding of the nature of our ethnographic encounters and our findings. We follow this by a description of our methods, participants and analysis.

In a graduate class on online communities taught by the third author, students complete a qualitative study of an online site using a combination of participant observation and interviewing. A project team of three students, including the first and second authors, chose to study Kotaku in Action. Not long after they began to request interviews on the site (following ethical guidelines (Bruckman, 2006)), this message was posted on KIA:

> "Dear KiA, if you are contacted by /u/gatech012 for this project, please be aware that this [is] indeed a trap, because the person doing the data collection and interpretation is intrinsically ideologically opposed to everything that this sub stands for."

In response to this, the third author replied, and volunteered (in Reddit tradition) to do an "AMA" ("ask me anything" discussion thread) with KiA users. The AMA took place on Feb 27, 2016 and includes 262 comments (Bruckman (submitter), 2016). This is when the community began to tolerate our presence, and started providing us valuable information. Following the AMA, many KiA members volunteered to speak with the research team.

Our study was approved by the Institutional Review Board (IRB) of the Georgia Institute of Technology. In all, we conducted thirteen semi-structured interviews with KiA users. All the interviews were conducted in

---

[2] Reddit members use /u/username to identify a Reddit user by his/her username.

Spring 2016. We recruited participants through private messages on Reddit. Participation was voluntary, and no incentives were offered for participation.

The interviews generally lasted between 60 to 90 minutes. Participants were asked questions about how they came to use the subreddit, what motivated them to continue posting on the subreddit, and their views on online harassment and moderation on websites for discussions on GamerGate. We conducted interviews over the phone, on Skype, and through chat. Some participants were contacted for brief follow-up interviews, for further clarification. We also shared an early draft of the paper with all the participants and they were given a chance to respond to it.

We read online postings of our participants, and compared their attitudes and actions to what they stated in their interviews. These were found to be largely consistent. However, because of the pseudonymous nature of Reddit platform, we constantly faced the possibility that our participants were lying to us. Following Phillips' approach (Phillips, 2015), we decided to note how the KiA users chose to present themselves to us, and deduce meaning from their (possibly choreographed) performance. Therefore, we present our findings as subjective perspectives and narratives of KiA rather than as objective facts.

Becker and Geer argue that any social group has a distinctive culture and a set of common understandings that find their expression in a language "whose nuances are peculiar to that group and fully understood only by its members" (Becker and Geer, 1957). They suggest to fieldworkers that "both care and imagination must be used in making sure of meanings, for the cultural esoterica of a group may hide behind ordinary language used in special ways." In studying KiA community, we paid particular attention to the terminology used by its members during our observations and interviews, and examined it as a function of their assumptions and purposes (Taylor and Bogdan, 1998). For instance, we have analyzed what terms like "political correctness" and "sealioning" mean to this community.

### *Participants*

Eleven participants in the study reported ages between 24 and 37. Two participants did not share their age, but one of them mentioned that he is in his 20s. Twelve participants were male and one was female. Two of the participants were from Norway, and the rest from the US. Four of the participants chose not to share all the demographic information we asked for. The interviewees included one current and one past KiA moderator. Table 1 shows some demographic information about our participants. We use light disguise (Bruckman, 2002) in describing our findings. Therefore, although we have omitted sensitive details to protect the identity of our participants, some active KiA active members may be able to guess who is being discussed.

Table 1: Study Participants

| ID | AGE | COUNTRY | OCCUPATION | INTERVIEW MEDIUM | GENDER | ACCOUNT CREATION | IS MODERAT-OR? |
|---|---|---|---|---|---|---|---|
| P1 | unknown | USA | Graphic Designer | chat | Male | Jun, 2015 | No |
| P2 | 26 | USA | Tech content manager | Skype | Male | May, 2015 | No |
| P3 | 27 | Norway | unknown | Skype | Male | Jan, 2015 | No |
| P4 | 20's | USA | unknown | Skype | Male | Oct, 2009 | No |
| P5 | 24 | USA | unknown | Skype | Male | Mar, 2010 | No |
| P6 | 24 | USA | Grocery-store manager | Skype | Male | Oct, 2014 | No |
| P7 | 25 | USA | Chemist | Skype | Male | Nov, 2014 | No |
| P8 | 28 | Norway | Computer Science student | chat | Male | Oct, 2013 | Current moderator |
| P9 | 34 | USA | Computer Programmer | chat | Male | Feb, 2013 | No |
| P10 | 37 | USA | Computer Technician | Phone | Male | May, 2014 | No |
| P11 | 25 | USA | Instructional aid | chat | Female | Feb, 2015 | No |
| P12 | 24 | USA | Student | chat | Male | May, 2012 | Past moderator |
| P13 | 26 | USA | Medical Professional | Skype | Male | Mar, 2011 | No |

Participants were self-selected—we spoke to members who were willing to be interviewed by a team of academics. Thus, our informants are likely some of the most moderate members of the group and are more engaged in talking about the topics at hand. They were eager to convince us and the world that they are reasonable people. We see specific evidence for this fact by comparing what some of them said during their interviews to their often more raucous and rude public presences on Twitter. As a result, in our dataset, the most moderate people are on their best behavior. However, their views are nevertheless revealing, and speak to broader issues.

*Analysis*

We fully transcribed data from the interviews and undertook multiple readings of these data. The first author used Dedoose software (www.dedoose.com) to perform coding, which underwent multiple iterations. Next, we performed inductive thematic analysis on these data and identified relevant themes and sub-themes through observation and discussion (Braun and Clarke, 2006). All the authors met regularly to discuss the codes and emerging themes, going back and forth between different categories and further scrutinizing data.

We collectively conducted over 100 hours of participant observation of KiA in an attempt to systematically understand the dynamics of the community. This was carried out between January-May 2016, and included lurking, commenting, voting on content, and posting content on the subreddit. We supplemented our interview responses with field notes and qualitative analysis of posts. We also corresponded with the current KiA moderators, who answered our questions through private messages, and helped us improve our understanding of the community. Over the course of this research, each of our stances developed as a result of engagement with the community, and through our evolving understanding of its objectives and characteristics.

We employed a mixed-method design for this study. In addition to the traditional ethnographic-style methods for observing online communities discussed above, we also used quantitative methods for analyzing user behavior on KiA. We assembled a sample of 1000 random submissions on KiA. We used PRAW, a python package that provides access to Reddit's API (PRAW, 2016), to gather these submissions. KiA encourages users to tag their posts using flairs. These flairs indicate the topic of the submission (Table 3). We extracted the flairs used in the submissions we collected to find the most popular topics of discussion. We also analyzed the website domains of these submissions that were links to external websites. To find what other issues KiA users are interested in, we extracted the set of users who posted these submissions. Following this, we collected all the postings made by these users on Reddit, again using PRAW.

Exactly what GamerGate is about can be hard for outsiders to understand. Before we move to our findings, we present a case study that describes the kind of arguments that individuals on the opposite sides of GamerGate debate make.

## GamerGate example : the Baldur's Gate controversy

Did gamers give negative reviews to the Baldur's Gate expansion because it is a bad game or because it featured a transgender character?

This question lay at the center of Baldur's Gate controversy. On March 31 2016, the 18-year old video game Baldur's Gate received a new expansion titled "Siege of Dragonspear" (BeamDog, 2016). The expansion received a barrage of negative user reviews on game shops like GOG and Steam.

Some reviews focused on problems with the game functionality like in-game bugs, dysfunctional multiplayer and mod incompatibility. Other reviews pointed out that the writing of the game was not up to the standard of the original Baldur's Gate games. Many reviews were also politically charged and criticized the inclusion of a transgender character in the expansion. They complained that the developers had crippled the game's creative strengths by shoehorning a token minority character to push their social justice agenda (Monroe, 2016).

Following this, Beamdog, the studio that developed this new expansion of Baldur's Gate, censored discussions about the expansion on its website (Imgur, 2016) and the official Steam forums. This resulted in a backlash from the gamers. They accused Beamdog of trying to cover up the problems in the game by attributing its poor reviews to extremist gamers. Further, many users on KiA were disappointed to find that the game taunted GamerGate by having one of the characters say: "Reeeeaaally, it's all about ethics in heroic adventuring" (Church, 2016).

Many users on KiA felt that the writers should not have hijacked the franchise by forcing their politics into the game. Some users claimed that they appreciate diversity of characters in games, but did not consider the inclusion of LGBT characters a special or revolutionary idea. The only thing that mattered to them is that the characters are well-written. One user posted on KiA:

"I'm trans and what pisses me off is the way the game does this. The trans character, when talked to, starts speechifying about gender before you're allowed to do anything else. Then, when you're finished, the only two options for reply are both positive and polite, which is incredibly immersion-destroying and completely against the philosophy of D&D/Infinity Engine games. In a game where you can murder almost anyone and everyone, and there's a dialogue option for most alignments, apparently not being nice to trans people would be a step too far. Story taking a back seat to politics."

Users on the other side of the controversy stated that the outrage was disproportionate to the perceived offense. They accused the GamerGate supporters of being transphobic, and believed that it was hypocritical of them to force game developers to remove the transgender character, since GamerGate opposes censorship of games. A user on GamerGhazi, a popular subreddit that hosts anti-GamerGate discussions commented:

*I'm very sad that it is even possible for human beings to carry that much hate. And I am even more saddened that a flourishing artistic medium for expression, like videogames, has become the place for those shitstain[s] to carry their battle against minorities. Fuck them.*

Beamdog released a statement that stated that "some of the negative feedback has focused not on *Siege of Dragonspear* but on individual developers at Beamdog, to the point of online threats and harassment" (Campbell, 2016). On the other hand, GamerGate supporters noted that the media eagerly covered the story from one side only, and portrayed them as an angry mob that harassed the progressive game developers.

In summary, there is no single explanation of the controversy about Siege of Dragonspear. Evidence supports the fact that it is indeed a badly designed game (Monroe, 2016). Many GamerGate supporters claim that the addition of the dialog line about GamerGate is proof that the developers were deliberately trying to provoke GamerGate supporters to get publicity. On the other hand, some vocal opponents of the game are clearly not supporters of transgender rights. This kind of complexity pervades our dataset.

## Findings

### KiA Community

In this section, we describe members of Reddit who interact with the KiA subreddit. Many of these users identify as gamers. They comment, post or lurk on KiA's discussion board or moderate it. They use the subreddit to share news relevant to GamerGate, and engage in discussions with one another.

We begin with a discussion of the topics that KiA members are interested in. Next, we discuss the demographic diversity of KiA. This provides context for understanding the beliefs and goals of KiA members, which we examine in the next subsection. Next, we describe the practice of archiving employed by the community, a practice crucial to understanding the ethos of the community. Finally, we analyze the conventions and policies that guide the activity of KiA users. This analysis helps explain how outside reactions to the community have shaped its perspectives. The description of the community in this section provides context for the members' views on harassment and free speech in the next section.

### User Interests

We used the Reddit data collected using PRAW to find the subreddits that were popular among KiA users. The ten subreddits where the highest number of postings were made by these users are shown in Table 2. The popularity of subreddits like "MensRights," "The_Donald" (a subreddit for discussing Donald Trump[3]) and "politics" shows that many KiA users are interested in political and social issues that go beyond gaming culture. The subreddits that provide informative gaming content and discussions like "pcmasterrace" (a subreddit for PC gaming enthusiasts) and "Games" are also popular.

Table 2: Subreddits used by a sample of KiA submitters

| SUBREDDIT | NUMBER OF POSTINGS |
|---|---|
| AskReddit | 6072 |
| MensRights | 6047 |
| pcmasterrace | 5843 |
| TumblrInAction | 5126 |
| The_Donald | 4021 |
| worldnews | 3535 |
| politics | 3415 |
| news | 3223 |
| GGFreeForAll | 3064 |
| Games | 2943 |

We analyzed the flairs that KiA submissions were tagged with. Table 3 shows the results of this analysis. The predominance of flairs such as "Opinion" and "SocJus" reflects a focus on events and issues that are misreported or under-reported in the mainstream mass media or gaming media in the community's view. Each time an event occurs that violates the principles of the community, the KiA platform works to report it, attract public condemnation of the event, and fuel subsequent action.

We also analyzed the website domains of KiA submissions and found that 73.3 percent of submissions with links pointed to YouTube, Twitter and Imgur. Very few submissions pointed to news sites like New York Times and Washington Post. This shows the community's reliance on more informal channels of news.

*Demographic Diversity*

It is difficult to empirically determine the demographic diversity on KiA because of the pseudonymous nature of Reddit. However, through our observations, we identified some basic demographic indicators. The ethos of the community is androcentric. Many users appear to have a libertarian attitude to society and culture. The threads on KiA often engage in discussions about American culture, media and politics, which indicates that a large number of users may be Americans.

---

[3] Data was collected during the 2016 Presidential campaign in the US.

Table 3: Flair Distribution of a sample of KiA submitters

| FLAIR | DESCRIPTION | COUNT |
| --- | --- | --- |
| Opinion | Opinion pieces by mainstream media outlets or individuals, both positive or negative. | 80 |
| Humor | Jokes, memes, parody articles, etc. | 74 |
| SocJus | Relating to social justice affecting wider nerd culture. | 67 |
| Discussion | Serious discussion on a topic or question. | 50 |
| Ethics | Major findings on unethical press behavior. | 42 |
| Censorship | Censorship of GamerGate discussion or of other gaming-related issues. | 31 |
| Misc. | Anything else with a tangential relevance to GamerGate. | 29 |
| Industry | Relating to the wider games industry and issues within. | 20 |
| Drama | News of personal conflict between individuals, often GamerGate key players. | 15 |
| Meta | Relating to internal KotakuInAction affairs. | 14 |
| Meetups | Relating to offline GamerGate meetups. | 12 |

Brad Glasgow conducted a survey of 725 individuals who support GamerGate on Reddit, Twitter or Voat (Glasgow, n.d.). He found that 89.2 percent of the respondents were male. 25.4 percent were between 16-25 years old, 55.4 percent were 26-35 years and 19.2 percent were more than 35 years old. 74.5 percent of the respondents were white, and 8.8 percent were of Hispanic or Latino origin. 52.7 percent were Americans. Although we did not conduct or find a survey like this on KiA specifically, these results provide some insights into the demographics of KiA.

*Goals and Beliefs*

KiA provides a platform to its members where they can share and discuss information. It also provides a space where they can organize and mobilize supporters to take actions for addressing issues like media cronyism and censorship in games.

KiA users have diverse opinions. They often have different perspectives on how the community should operate. In mid-2015, the community split over the argument of what its focus should be. Some users believed that KiA should remain focused on its original goal of improving the standards of ethics in gaming media[4]. Many others argued that PC culture and third-wave feminism were responsible for many problems that pervade not just the gaming industry, but also the society more broadly. They insisted that the community should expand its objectives to fight against PC culture. In early 2015, the latter argument won, and the community expanded its focus. However, many members, including P12, a prominent KiA moderator at the time, left the community in protest.

In early 2017, after a community discussion, the moderators added a "points system" for new posts, to focus more strongly on core issues about gaming and nerd culture (see Figure 1) (ITSigno (submitter), 2017). A post must have two points or it will be removed by the moderators. "Related politics" is defined as relating to the Internet, gaming, and censorship. The exact details of the points system remain a topic of active debate, with changes made by votes of members.

Based on our interviews and observations, the community highly values freedom of expression, and opposes censorship in games and online communication. Many KiA members see themselves as average people and they consider GamerGate a grassroots movement to share their opinions and information about inept gaming companies and to bring about a change in the gaming industry and media.

**POSTING GUIDELINES**

| Feature | Points |
|---|---|
| Gaming/Nerd Culture | +2 |
| Journalism Ethics | +2 |
| Official Socjus | +1 |
| Campus Activities | +1 |
| Related Politics | +1 |
| Censorship | +1 |
| Media Meta | +1 |
| OC Artwork | +1 |
| Socjus attack by *media* | +1 |
| Unrelated Politics | -2 |
| Memes | -2 |

Posts that have less than 2 points will be removed.
Self-posts with a reasonable argument establishing relevance/importance can bypass the posting guidelines (but not other rules)

Figure 1: KiA Posting Guidelines

The community also holds that the political correctness has gone too far in our society. KiA's conception of political correctness can be described as what they consider a relentless push by SJWs to politicize every aspect of our society (especially gaming culture) and to accuse their opponents of holding views that could only be motivated by misogyny and bigotry. KiA users believe that they should actively work to fight against this PC culture. Participant P11 said that the community values encouraging an environment where people don't "feel like they need to walk on pins and needles to avoid offending others." Participant P3 said:

> "What KiA stands for is that censorship in general, in art or if it's a person - that should not be censored just because you get offended and, of course, within the law. So, if it's an art that gets censored like games - that's something people usually are very against" (P3)

---

[4] KiA's idea of "ethical media" constitutes games journalism that ensures disclosure of relationships between game developers and reviewers, if any, and guarantees that games are reviewed according to the principles of journalistic objectivity (Meditations, 2016). Some KiA users assume that any injection of politics in game reviews is unethical whereas others consider it appropriate if the subjective biases of the reviewer are disclosed.

Members believe it is their duty to oppose "social justice warriors" (SJWs). In the KiA world, an SJW is "*someone who has gone completely off the deep end in terms of ideology and logical fallacies. It's us versus them. They're always right no matter what*" (P13). Many KiA members accuse SJWs of being self serving, but insist that their own activism is selfless and legitimate. Participant P10 said that it is impossible to disagree with an SJW in any way on any topic, because, "*As soon as you disagree with them, you are a misogynist, you are a racist...you hate women, you are a rape apologist.*" Members of KiA believe that they are often falsely accused of a panoply of anti-social beliefs and actions. While the term "GamerGate" is synonymous with "hate group" to many outsiders, to KiA members, it is a positive affiliation to a group with values that challenge the status quo in a constructive way.

The KiA community tends to respond with anger to anything they see as attacking gamers. However, there is some disagreement about what constitutes legitimate criticism. One KiA poster writes about criticism of games, "*Some opinions are really against [all] video games and they're stupid. But some opinions are legitimate criticism of trends in game design. 'I'd like more non-violent games' is a legitimate opinion. But how to tell the difference*" (KotakuInAction, 2015)? What constitutes fair criticism (either of them or by them) is a controversial issue for the community.

Many participants stated that they use KiA to receive news on GamerGate and social justice issues. Some noted that they use KiA to organize "real-life" meetups with other KiA members. Many members believe that the media is unfair to the community, and KiA serves as an alternate news source by providing content that they do not find on traditional media. Other KiA users believe that the community serves as a "media watchdog."

> "It's kind of, just sitting and waiting for crap to happen. Just to watch and call out, when media is acting up, when they are lying and, you know, pointing out the nepotism." (P10)

*Archiving*

The community believes that record-keeping is important, and uses a number of tools to preserve records. One of the rules of KiA recommends its members to "archive as many things as possible." Archiving preserves articles in their original format, so that alteration or disappearance of embarrassing records from websites can be exposed, and media can be held accountable. 'Archive.is', a website that allows its users to save a text and a graphical copy for any webpage, is used for this purpose. One of the KiA users created "mnemosyne," a bot that automatically archives each submission on KiA.

The community also has an active blacklist of websites, and a bot automatically filters postings of submissions that link to any of these sites for review by the moderators before being posted. The submitter is also notified that this review can be bypassed, if the article is archived and resubmitted using the archived link. Members believe that the websites on this blacklist feature articles with sensationalist headlines related to GamerGate, so that they attract visits, and generate online advertising revenue. By using archiving, the community denies "click revenue" to these websites. Although KiA is firmly against censorship, the use of such practices on KiA indicates that members view their own moderation practices as quite different from those by others.

Each time the moderators take any action, a bot automatically posts the decision and a link to the KiA page where the decision occurred to a feed on modlog.github.io, an external website dedicated to building feeds of Reddit moderation. The deleted links are also tweeted on a public Twitter handle "@KIADeletedLinks." Though we have no detailed accounts of why this practice was adopted, the Twitter handle mentions "KiA's Transparency Pledge." We suspect that this practice represents the political value of transparency in governance.

*Conventions and Policies*

The activity on KiA is guided by (1) a set of established rules defined by Reddit guidelines; (2) a set of emergent rules that are specific to KiA; and (3) the norms of KiA. We discuss each of these in this section.

<u>Conventions and Policies: Reddit's Content Policy</u>
Among other rules, Reddit's content policy dictates that users are not allowed to post content that "threatens, harasses or bullies or encourages others to do so" (Reddit.com, 2016). The policy also describes how its rules are enforced, and ways of enforcing include banning of Reddit communities. KiA, like all subreddits, is governed by this policy. Some participants said that the biggest concern of KiA moderators is that the subreddit would get banned under this policy.

> Outside of our little niche, there's a lot of misinformation and general just dislike of us, and the admins are probably always looking for good reasons to ban us. (P6)

The policy also states that individual subreddits may have their own additional rules. One instance where such local rules are enforced is when some subreddits like "Rape" and "BlackHair" ban any accounts that post submissions or comments on KiA (Figure 2). One member said that there are other subreddits (e.g., r/GamerGhazi) where a user gets banned if he claims his support for GamerGate. Some participants stated that they considered these policies unreasonable.

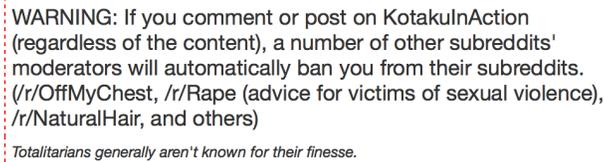

Figure 2: KiA warns users when they post or comment

### Conventions and Policies: KiA Rules

KiA has its own additional rules. The sidebar of KiA highlights these 9 rules[5]: "(1) Don't be a dickwolf[6]; (2) No "Personal Information"; (3) No Politics; (4) Please tag posts for flair; (5) We are not your personal army; (6) Archive as much as you can; (7) Don't post bullshit; (8) No Reposts; (9) No MetaReddit Posts."

One KiA user interpreted the rule "Don't be a dickwolf" like this: "It means say whatever you want, but don't start hurling insults at fellow KiA members when you don't like what they say." The rule "We are not your personal army" forbids brigading[7], dogpiling (described under 'KiA and Online Harassment' section) and creating call-to-arms posts against individuals. Links to comments of other subreddits are automatically banned by a bot. The "Don't post bullshit" rule prohibits users from posting

---

[5] These were the rules on KiA at the time of data collection. These rules continue to evolve over time.

[6] The word "Dickwolf" originated in a controversy over a 2010 comic strip (Fudge, 2013).

[7] Brigading is a concerted attack by one online group on another group, often using mass-commenting.

"editorialized headlines and links to provably false information" (Kotaku In Action, 2016). It urges members to provide information without trying to spin a narrative.

Conventions and Policies: KiA Norms

The norms of the community dictate that moderation should be minimal. Many members strongly believe in freedom of expression. There is a spectrum of interpretations of limits on free speech, and KiA leans towards favoring no limits. Many participants expressed their concerns about social media platforms shutting down parts of the political spectrum by censoring selected conversations or banning certain users. The community claims that it values discussion, and it believes that everyone should be allowed to have an equal voice. Members consider that this belief guides the norms of the community, where users are not banned if they express an unpopular opinion.

> "I think the correct term would be "laissez-fair," the kind of hands-off moderation that allows us to really post anything that is related to our pretty lofty generalized goal. So you can get any movement going as long as it's kind of related, is one big benefit." (P6)

However, the voting mechanism of Reddit does not allow posts with unpopular viewpoints to appear at the top. Some participants admitted that they don't often see anti-GamerGate posters on KiA, and even when such users show up, their posts frequently get down-voted, and thereby buried under the more popular pro-GamerGate posts.

The moderators try to find a good balance between freedom of speech and on-topic discussions. There are ways to get banned or to have a post deleted, but such penalties are not likely the result of using unacceptable terminology or expressing an unpopular point of view.

Even though the moderation is minimal, it still has a significant impact. For example, Participant P13 posts content on Twitter that he says he is sure would never be allowed on KiA.

The community embraces a few norms and practices that are widely condemned by outsiders, and some of these norms have emerged in response to the reaction of outsiders to the community. Consider that the primary visual draw on the KiA subreddit header is the image (Figure 3) of a red-haired woman named Vivian James, a frequent character in GamerGate-related comics. This image depicts Vivian riding a sea lion while wearing a sock-puppet on one hand. This reference two separate practices that are important to the community:

Figure 3: KiA Header Image

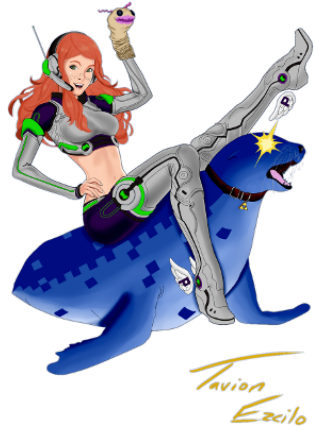

(1) "Sock-puppeting": This refers to the creation of auxiliary accounts by a user to provide anonymous support to arguments posted by his main account. Anti-KiA communities often blame KiA for engaging in identity deception using sock-puppeting. KiA strongly refutes such allegations, and uses a sock-puppet in its header image to poke fun at them.

(2) "Sealioning": This refers to the act of persistently but politely requesting evidence in a conversation. The name "sealioning" comes from a comic in which a sea lion annoys a couple by trying to engage them in a conversation that they are not interested in having (Malki, 2014). Sealioning is viewed by anti-GamerGate users as intrusive attempts at engaging an unwilling debate opponent and excessively requesting evidence. However, KiA has cultivated and embraced "sealioning" as a rhetorical norm in which members practice providing and requiring evidence and source material while engaging in regular conversations. As we will discuss later, the presence of trolling might have encouraged this practice too. A notable example of collecting evidence is the frequent use of the site deepfreeze.it, on which KiA users and other GamerGate supporters document evidence of unethical gaming journalist behavior for use in later arguments.

### *KiA and Online Harassment*

We began this research with a set of questions about online harassment. Do KiA members engage in harassment? What do they think "harassment" is? In this section, we discuss participants' response to accusations of harassment.

We asked many of our interview subjects what they thought "harassment" is. We also analyzed discussions about online harassment on KiA. A consensus for the definition and resilient characteristics of harassment from the community's perspective emerged from the collated responses. Participants divided communicative acts by their intensity.

Intensity has two components: the content of an individual message, and the frequency with which messages are sent.

*Content*

Participants felt that a single message may or may not be harassment, depending on the content:

> "I think that a big problem is the redefinition of what harassment means. I think…just sending mean tweets is considered harassment, as opposed to just like telling someone to explicitly kill yourself… there's definitely different degrees." (P2)

> "I think, GamerGate sees harassment in the same way that any average, ordinary person or even the law views harassment - which is you know, stalking, death threats, as in calling your house; a very persistent, when people are persistent in their stalking, or harassment, basically the way the law sees it." (P10)

This indicates that in some instances, KiA members may not grasp the emotional toll that their remarks can exert on other users. As Whitney Phillips describes, "even the most ephemeral antagonistic behaviors can be devastating to the target, and can linger in a person's mind long after the computer is powered down" (Phillips, 2015).

Many participants expressed their concerns about conflation of criticism with harassment. Some claimed that opponents of GamerGate use accusations of harassment to fight against differences of opinions. They felt that a critical consequence of this conflation is affective desensitization of many users in the community to the concept of harassment.

> "It is ultimately derailing 'cause the entire thing has been, we're trying to have one conversation and the other people just call us sexists and harassers, and it's like, "that's not a response."" (P2)

> "The accusations, the racist and misogynist, that is their one go-to insult to shut down anything that you have to say." (P10)

Some participants said that KiA opponents sometimes consider expressions of sincere disagreements with them as harassment and block them.

> "You could be in a debate with somebody else, and you could be asking for what are the gun statistics from 1980, and they'll be like "That's a sealioning question. I don't trust you. You're blocked" …and I'd be like, what the fuck just happened here?" (P13)

*Frequency*

Some participants felt that postings by a sole user on a single social network should not be considered harassment, since almost every social network provides its users the ability to 'block' accounts. When a user blocks an account, the blocked account can no longer send messages to the user. Participants argued that harassment entails a more persistent behavior, where the harasser is willing to create new accounts when banned or blocked, and continues to send threats to the victim.

> "Harassment requires a little more motivation, a little more intent, a little more longevity. Someone kind of drunkenly wandering up to your house in the middle of the night and knocking on your window isn't stalking. But doing it 20 times might be." (P6)

Another scenario is that of "dogpiling." KiA rules explicitly prohibit calls for dogpiling, which is an indication that the community has had previous trouble with dogpiling. Dogpiling occurs when a single individual is overwhelmed by receiving a large number of messages from different people. Such messages are often sent through private channels on social networks, and therefore it is not always obvious to the sender that the target is receiving hundreds of similar messages. Although the content of individual messages may not be seriously threatening, the receiver feels vulnerable and threatened.

*Justification*

Many members also distinguished among different communicative acts by whether they considered the act to be an appropriate response to a prior offense. Some participants observed that individuals who commit social transgressions in public deserve to be called out for their actions.

> "You're in a public space. If you're acting like a child, if somebody films you, that's kind of your fault." (P13)

A few felt that such actions also include the postings made on websites like Twitter. They argued that such websites should be considered public spaces because the content on them is publicly available.

Many members also took into consideration the status of the account receiving attacks. They felt that individuals who are "public figures" should expect to lose certain protections. Participant P13 distinguished between messages sent to identifiable, personal accounts versus those sent to anonymous or organization accounts.

A few participants discredited harassment claims, and stated that harassment victims were soliciting their own abuse. Participant P2 noted that some people provoke angry responses by saying something inflammatory, and then present responses as evidence of harassment.

> "My personal opinion is that it seems like there is a weird culture of victimhood, where victimhood is put up on some kind of weird pedestal." (P4)

*Accountability*

Some participants felt that in a large, leaderless community like KiA, it is difficult to rein in every user, and problems like harassment are bound to occur. They assume that the nature of the Internet and online social networks makes harassment in any contentious online movement inevitable.

> "I think this is the biggest thing the media fails to get about Gamergate: it's basically the same as the rest of the Internet outside Gamergate. The Internet is an incredibly hostile place sometimes." (P9)

A few participants admitted that some users who subscribe to KiA have engaged in harassment. However, they claimed that such users do not represent the values of KiA and are at the periphery of the community. They challenged the accusation that organized harassment exists in the community.

> "Do I think that there are still elements of KiA who support harassment? Yeah, of course. I actually had an argument the other day with someone who wanted to get somebody fired." (P5)

*Perceptions of Unfair Portrayal*

Many participants felt that it was unfair of the media and members of other online communities to label everyone associated with KiA as harassers. They believe that an overwhelming majority of KiA members act responsibly and do not cause any problems. In their view, most of the users in the community strongly condemn harassment and doxing, and many work hard to ensure that such activities do not occur from inside the community. One former KiA moderator said:

> "Reading what was being said about KiA, that it was a front for an abuse campaign, was enraging. The mod team had done all we could to keep that sort of stuff out of the sub, and to discourage it at every possible avenue." (P12)

Many participants said that a number of users in KiA also got harassed and received death and rape threats, but such incidents were dismissed by the media and KiA opponents. In their view, the media deliberately under-reported such incidents to spin the narrative of KiA as a hate group, and undermine its arguments about ethics in game journalism.

> "Let's not forget that there is a lot of talk about harassment of these female developers and these prominent people, but there are a lot of people who supported KiA who were harassed, who were doxed and who were kind of blacklisted because of their support for this, basically this idea." (P5)

*Doxing*

KiA's 'No "Personal Information" rule' bars users from sharing individuals' phone numbers, addresses, and other private information, and asks them to avoid posting links into people's Facebook pages. A few participants stated that they haven't seen any incidents of doxing on KiA. Others mentioned that they have rarely seen doxing, but ignored such posts.

> "Most people don't care. And so somebody says like this is a person I hate and this is his home address, you don't care. You would look away. You would be slightly disgusted. That's how most people felt, so they didn't look at it." (P4)

Participant P8 claimed that doxing is likely to occur only to users who are known over several platforms, and are "something of a public figure." It was common for many participants to state that the claims of harassment and doxing were largely exaggerated.

Participant P4 investigated some incidents of doxing, and found that they were being organized by a troll group. Some participants pointed out incidents where KiA users got doxed. Participant P12 stated that his personal information was revealed, but the source of doxing information was promptly suspended, and the information was quickly removed.

Some participants claimed that they have seen information presented as evidence of doxing on KiA that they wouldn't consider doxing. This raises questions about the boundaries of what should be considered doxing. For instance, if information is readily available using an Internet search engine, should it be considered doxing? One moderator explained how KiA deals with incidents of doxing:

> "We have a pretty limited set of tools for when that happens. Naturally - if people post dox on KiA, we remove it. If someone decides to post personal information on another site, there's very little we can do." (P8)

A few participants pointed out the existence of a "harassment patrol" in the community's early days. It consisted of a group of users who actively looked out for trolls on the community, and took actions to ban them. They also reported doxing incidents and prevented users from organizing brigading activity on the community.

*Trolling*

KiA hosts a variety of fast-moving discourse that includes good-natured ironic posts, humorous or sarcastic comments, pranking and sensationalist exaggeration. The more antisocial aspect of this discourse is trolling. Trolling entails provoking others to engage in pointless, time-consuming discussions (Herring et al., 2002; Kraut and Resnick, 2012; Donath, 1999).

In her study on Internet trolls (Phillips, 2015), Whitney Phillips notes, "Trolls believe that nothing should be taken seriously, and therefore regard public displays of sentimentality, political conviction, and/ or ideological rigidity as a call to trolling arms." The strong ideological stances on both sides of the GamerGate controversy make it an attractive target for trolls. They often disrupt the discussion space on KiA and other GamerGate-related forums. Some participants also accused such users of "false flagging" (falsely blaming KiA users for operations that they did not conduct).

A few participants talked about troll groups, some of which are splinter groups that emerged out of the GamerGate forums on 4chan and 8chan websites, that attempt to disrupt and mislead discussions about GamerGate. Some "third-party trolls" harass members of both GamerGate and anti-GamerGate communities, and blame the other group, to instigate the groups to fight each other. GamerGate supporters have often claimed that "much of the mayhem associated with the movement comes from third-party trolls who get a kick out of baiting both sides" (Young, 2015).

A few KiA users have called for separation from #GamerGate and rebranding under a new tag to distance the movement from its association with harassment in the popular zeitgeist. Such calls were overwhelmingly rejected by the community because it felt that any rebranding would divide the community, and the trolls would simply follow the movement and smear it under the new tag.

The presence of these trolls might have affected the discourse on KiA. For instance, users are often asked to provide evidence to back up their claims, so as to ensure that they are not trolling. The KiA rules also prohibit posts and comments that "are clearly not intended to generate discussion, but rather just aimed at generating as much drama and outrage as possible" (Kotaku In Action, 2016).

*Legal Discourse*

Some participants mentioned that instances of harassment and legitimate harm should be taken seriously, and such instances should be reported to and handled by authorities.

> *"If it's actual harassment, go to the police. That should be the end of it. But they make the argument that police don't do enough to help, and that's probably true. But, they probably have more pressing things than somebody bothering someone on the Internet." (P2)*

Participant P2 expressed concern that illegitimate cases of online harassment might discourage the authorities to deal with legitimate cases in the future.

All of our participants said that they do not personally engage in harassing others, and that KiA as a group specifically prohibits all forms of harassment. This is in sharp contrast to the portrayal of GamerGate in the popular press. There are a number of possible explanations for this apparent contradiction. First, there is a self-selection bias in how our interview subjects were selected. The people who would agree to speak to academics about KiA are likely not the ones doing the harassment. Second, because KiA is a large, leaderless group, it is impossible to hold the group accountable for the behavior of any single individual, because that individual is simply redefined as not speaking for the group (like the "no true Scotsman" logical fallacy (Fieser and Dowden, 1995)). Third, underlying this contradiction is a sincere disagreement about what is "harassment" and what is free speech. A fourth potential explanation is that our informants are simply lying to us. Although they are clearly putting their best selves forward, we believe them to on the whole be giving honest accounts. The other three explanations are all true in varying degrees. We conclude that KiA is not a viper's nest, though there are probably vipers in the nest.

## Discussion

As we have seen, KiA members believe that they have been wrongly accused of harassment. The other side of the controversy, of course, has a radically different account of what has actually taken place. It is indisputable that both sides have both experienced and committed harassment. What is impossible empirically for us to determine is the relative prevalence of harassers in the community. Our informants state that a few bad apples are giving the entire group a bad name. Others label KiA a hate group and insist that it is immaterial that there are a small number of decent people mixed in. The relative prevalence is an empirical question that we can't answer, and nothing in this paper should be construed to support one view or the other.

In her book "Hate crimes: Causes, controls and controversies," Phyllis Gerstenfeld asserts that there is no simple way to define a hate group, and "whether a particular group is to be classified as a hate group is often in the eyes of the beholder" (Gerstenfeld, 2013). She argues that one of the problems with identifying hate groups is that "some organizations have certain factions that are clearly bigoted although other factions are not."

When a movement is made up of people with differing views and tactics spread across multiple websites, it is impossible to hold the group accountable for the action of any individual, because the individual's actions can always be redefined as not representing the group. Trying to hold an entire loosely defined group responsible for the actions of its worst behaved members appears to be a catalyst for escalating rancor on all sides. Therefore, we should encourage efforts to understand the commonly held values of such groups instead of characterizing them by the views and actions of outliers.

### *Implications for Theory: Free Speech vs. Harassment*

Prior research has shown that anonymity provided by Reddit, Twitter and other social media websites lowers social inhibition, and encourages users to be more aggressive in their communications (Kraut and Resnick, 2012). This leads to situations in which some users see their online behavior as innocuous or an exercise in free speech, but it is construed as online harassment by others.

In her study on cyber-racism, Jessie Daniels notes that there is a US/Europe cyberhate divide (Daniels, 2009). She explains that the US response to white supremacy online is to view it as speech protected under the First Amendment and to forfeit it only when it is joined with conduct that threatens, harasses, harms, or incites illegality. In contrast, other Western industrialized democracies address online racism by broadening the scope of their existing antiracism laws.

There exist similar disagreements on the questions about the content of other hateful material. The enormous international influence of the US policies and its prominence as a safe haven for hosting Internet hate speech reduce the likelihood that nations who wish to regulate hate speech online will be able to do so. Besides, as Titley et. al point out, "there seems to be consensus that the problem of cyberhate is increasing both in magnitude, and in the variety of strategies used" (Titley et al., 2014). This reflects a need to consider analytic alternatives to the binary interpretation of free speech versus harassment that many KiA users seem to hold. Distinguishing controversial speech from hate speech, and weighing "freedom of speech" against protection from abuse would help the researchers and regulators think more critically about these issues.

These tradeoffs are further complicated by the presence of trolls who pretend to be sincere members of the community and lure others into pointless discussions. Therefore, efforts to characterize the evolution of troll behavior along the lines of Whitney Phillips' work on trolls (Phillips, 2015) should be encouraged so that platforms can efficiently identify trolls and regulate their postings.

Some researchers argue that the Internet has witnessed a number of moral panics regarding online activity, and this clouds the fact that only a small minority of users actually engage in disruptive or illegal activity (Ellison and Akdeniz, 1998). Our findings indicate that there may be an aspect of moral panic in response to GamerGate. Many KiA users believe that they experienced one-sided reporting by the media. They argue that misinformed reactions and stereotyping by their opponents fueled the anger of GamerGate supporters and intensified their activities. To break this cycle of negative reinforcement and escalation, our findings suggest that it is advisable to take a more balanced tone in response. If outsiders seek to find common ground with the more moderate members of the movement, this can break the cycle of escalation.

While outsiders might seek mutual understanding with the more moderate members of the group, it is clearly not ethical or strategic to appease the true harassers in any way. But how do we tell the difference?

Our findings suggest two key dimensions to understand controversial speech: its intensity, and whether it is perceived as justified from a particular perspective (see Table 4). Intensity has two dimensions: the strength of each utterance, and how often the communication is repeated.

Table 4: Conceptualizing Controversial Speech

|  | JUSTIFIED | UNJUSTIFIED |
|---|---|---|
| HIGH INTENSITY | Public Shaming | Harassment |
| LOW INTENSITY | Criticism | Insult |

A single utterance can qualify as "high intensity" if it is strongly worded or contains an actual threat. On the other hand, a simple utterance (like "you're wrong") might be perceived as intense if it is repeated many times. As we discussed in our findings, some users may feel harassed when they receive a large number of messages, even if the content of individual messages may not be seriously threatening from the sender's perspective. In such situations, it may not be reasonable to label all message senders as harassers, even though harassment has occurred.

When people feel that their intense criticism is justified, the activity is often called "public shaming" rather than harassment. As Jon Ronson has thoughtfully documented (Ronson, 2015), public shaming often has consequences for the individual (like loss of job) that outweigh the perceived offense.

Both intensity and justification are subjective. Underlying much of the controversy we observed are *disagreements about what quadrant we are in*. One person's "criticism" is another person's "harassment."

Judgments of intensity differ radically depending on an individual's basic views on the proper limits on free speech. Citron (Citron, 2014) writes that although online speech is crucial for self-government and cultural engagement, certain categories of low-value speech, e.g., true threats, defamation, fraud and obscenity, "can be regulated due to their propensity to bring about serious harms and slight contribution to free speech values." Everyone agrees that you can't yell "fire" in a crowded theater. Beyond concrete and immediate harm, where do we draw the line? Feminists and critical race theorists argue that words have power, and we are responsible for the emotional harm our words may cause others (Spender, 1985; Daniels, 2009). Strong civil libertarians argue that censorship is a slippery slope, and freedom of speech includes the right to offend (Brennan, 2012).

With such fundamental disagreements on what sort of speech is appropriate, it's a wonder that people ever succeed in civil communication. Fortunately, this problem is normally solved by social norms. Members of different online communities develop a sense of local social norms for appropriate communication. Each subreddit in fact evolves its own norms—what you may say on Kotaku in Action is quite different than what you can say on GamerGhazi or AskReddit or any of the other thousands of subreddits. Conflict about appropriate speech is particularly likely to emerge on sites like Twitter, where it is not clear whose social norms apply.

### *Implications for Design*

A key approach to managing the problem of online harassment is by developing moderation and blocking mechanisms (Crawford and Gillespie, 2016; Lampe and Resnick, 2004; Geiger, 2016). Our findings add nuance to our understanding of the challenges of this undertaking. As we discussed, the tradeoffs between online harassment and free speech are complex. Couching too broad a spectrum of online dispute under a single umbrella of harassment can lead to broad reactionary interventions that are problematic. Although it may appear that laying out detailed, formal rules to guide moderation would bring a sense of fairness in an online community, the moderators need enough flexibility to judge any action in its context. Moderation decisions should take into account the intensity of the language used, as well as the frequency of communications directed at a single target. As Phillips recommends, moderation decisions should also consider the persistence and relative searchability of data for a given behavior (Phillips, 2015). Supporting a personalized approach to controlling the user's social feed should also be encouraged.

We argue that another, complementary direction where designers can focus is the design of tools that can help improve discussions and mutual understanding of groups with different ideologies. This is a challenging problem. For example, consider the context of GamerGate. It is difficult to create a legitimate dialogue between the two sides: Basic language choices (for example, KiA's use of the phrase "social justice warriors") posit deep-seated assumptions about the other side. The opponents of GamerGate view it as a hate group, while its supporters believe that their legitimate concerns are rebuffed by portraying them as harassers. We propose that one useful way to address such challenges is to draw from prior research on modeling argumentation for the social semantic web. There is a vast body of work in this area where researchers have proposed theoretical models and implemented social web tools that help users engage in argumentative discussions (Schneider et al., 2013). For example, Kriplean et. al. developed ConsiderIt, a platform that allows users to author pro and con points (Kriplean et al., 2012). This augmented personal deliberation helps mitigate the opportunities for conflict that occur in direct discussions while allowing users to consider the arguments on the other side.

Differences in cultural norms among groups can make communications difficult. It would be helpful to design solutions that help bridge across different norms of politeness. Disentangling the mode of address from content can help. Grevet's work on designing social media to facilitate more civil conversations provides useful insights (Grevet, 2016). Such tools can help users identify common ground.

Going forward, we intend to explore other avenues of research in designing interfaces that help people understand one another. As a first step, we are currently studying ChangeMyView[8], a community that facilitates discussions between users with opposing viewpoints. We are analyzing how the design mechanisms and social norms of this community allow users to engage in meaningful conversations with people they disagree with. This project aims to explore how we can motivate such civil discussions on other platforms.

### *Limitations*

In this research, we deliberately sought out just one side of the controversy: Who are the people on KiA, and how do they view their activities? We are not attempting to make any statement in favor of or against members of KiA, but simply to try to see what the world looks like from their point of view. What we found had much more complexity and nuance than we originally anticipated. In the future, it would be interesting to study individuals who oppose GamerGate on Twitter and sites like GamerGhazi.

A key limitation on our findings is the nature of our sample. Our sample size is small but we triangulated our interview data with notes made during participant observation on KiA. We note a self-selection bias – we only spoke with KiA members who were willing to talk to us. Additionally, social desirability bias might have motivated our interviewees to under-report behaviors that may be viewed as unfavorable.

---

[8] www.reddit.com/r/changemyview

We hope that our study has been fair even though, given the size and diversity of the KiA community, there may well be important exceptions to what we have described and observed.

## Conclusion

What made Lindy West's story so compelling is that she and her harasser transcended their differences and reached a degree of mutual understanding. Nothing like that happened in this study. We have no reason to believe that anyone we spoke to developed any new insights into how their actions might affect others. We may perhaps have helped outsiders to develop some understanding of our subjects and their concerns. Members of KiA have a mix of concerns some of which a neutral outside observer might find reasonable to a degree (like frustration with political correctness, frustration with policies of the game industry, and concerns about the quality of journalism), and other concerns likely less so. However, dismissing their concerns entirely simply fuels their righteous anger. The path to greater harmony is through mutual understanding. As it was for West and her harasser, that understanding needs to be two-way. The intriguing question for the research community is whether it is possible to design tools and systems to help foster such understanding.

In this work, we have spoken with a self-selected subset of members of the KiA subreddit who were willing to speak with academics, and were likely on their best behavior while doing so. However, even from observing this tiny slice of the broader community, we found fascinating and unexpected complexity and nuance.

Clifford Geertz writes that anthropologists don't study villages—they study *in* villages (Geertz, 1973). Studying in this particular techno-village, we observed, first, a fundamental problem of accountability in distributed social movements. New tools that help visualize the beliefs of groups might help outsiders distinguish common beliefs from rare ones. Second, we find that what is "harassment" is often in dispute. Our informants complain that simple disagreement on their part is often portrayed as harassment. It is difficult to resolve issues of possible harassment if we cannot even agree on whether it is taking place. Different design solutions may be needed for addressing deliberate harassment versus sincere misunderstanding and communicating across different social norms of conversation. Although much work has been done on blocking tools for deliberate harassment, there are a host of open research questions about how to create tools to support greater understanding. Our findings suggest that barriers of language use and differences in social norms of politeness often obscure underlying common values, and these challenges may be amenable to designed solutions.

## About the Authors

**Shagun Jhaver** is a PhD student in the School of Interactive Computing at the Georgia Institute of Technology. His research aims to inform the design of efficient and equitable online moderation systems. His prior work has focused on understanding how social media users perceive opaque moderation systems and how these perceptions shape their behavior. Jhaver received his MS in Computer Science from the University of Texas at Dallas in 2014 and his B. Tech. in Electrical Engineering from Indian Institute of Technology Bombay in 2010.

**Larry Chan** is a user-experience researcher at Illumio. He completed his Master's degree in Human-Computer Interaction in the School of Interactive Computing at the Georgia Institute of Technology in December 2016.

**Amy Bruckman** is Professor and Associate Chair in the School of Interactive Computing at the Georgia Institute of Technology. Her research focuses on social computing, with interests in online collaboration, social movements, and online moderation.  Bruckman received her Ph.D. from the MIT Media Lab's Epistemology and Learning group in 1997, her M.S. from the Media Lab's Interactive Cinema Group in 1991, and a B.A. in physics from Harvard University in 1987. More information about her work is available at: http://www.cc.gatech.edu/~asb

## Acknowledgements

The authors would like to thank Ian Stewart, who conducted three of the interviews in this paper as part of his team project in CS 6470 Design of Online Communities at Georgia Tech. Eric Gilbert participated in numerous research meetings about this paper, and suggested the title. We thank the Social Computing researchers in the School of Interactive Computing at Georgia Tech for their valuable feedback in preparing this manuscript. We also thank the Kotaku in Action community for allowing us to study their social dynamics and understand their perspectives.